\documentclass[12pt]{article}
\usepackage[dvips]{graphicx}

\title{The use of the GARP genetic algorithm and internet grid computing in the Lifemapper world atlas of species biodiversity }
\author{ David R.B. Stockwell\thanks{Corresponding author: Tel: 1 858 8220942, Fax: 1 858 8223631, Email: davids@sdsc.edu. Address: San Diego Supercomputer Center, 9500 Gilman Dr, La Jolla, CA 92037.} 
\and James H. Beach, Aimee Stewart, Gregory Vorontsov, David Vieglais, and \thanks{Biodiversity Research Center, 1345 Jayhawk Boulevard, University of Kansas, Lawrence, Kansas 66045} 
\and Ricardo Scachetti Pereira \thanks{Centro de Referencia em Informacao Ambiental, Av. Dr. Romeu Tortima, 388, Campinas, Sao Paolo, Brazil 13084520 }}

\begin{document}

\maketitle

\begin{abstract}
Lifemapper (http://www.lifemapper.org) is a predictive electronic atlas of the Earth's biological biodiversity. 
Using a screensaver version of the GARP genetic algorithm for modeling species distributions,  
Lifemapper harnesses vast computing resources through 'volunteers'
PCs similar to SETI@home, to develop models of the distribution of the worlds fauna and flora.  
The Lifemapper project's primary goal is to provide an up to date
and comprehensive database of species maps and prediction models (i.e. a fauna and flora of the world) using available data on species' locations.  
The models are developed using specimen data from 
distributed museum collections and an archive of geospatial environmental 
correlates. A central server maintains a 
dynamic archive of species maps and models for research, outreach to the general community,
and feedback to museum data providers.  This paper is a case study in the role, use and justification of a genetic algorithm in development of large-scale environmental informatics infrastructure.


\end{abstract}

\section{Introduction}

Humans have explored the life of the planet for the past 250 years. That knowledge is documented by millions of original specimens of plants and animals in the world's natural history museums and herbaria. However, we have yet to achieve the goal of identifying and mapping the distribution of the species present on earth. The vision of the Lifemapper system was to use the Internet to retrieve records of species locations from museum collections, to compute the ecological profile of each species, and predict where each species could potentially live. This would then be the basis of a database of species distribution maps.  The approach of developing models of species' habitat provides not only maps of where Earth's species of plants and animals live, but also predictions of where Earth's species of plants and animals could potentially live under different scenarios such as climate change, and where and how introduced species could spread across different regions of the world. 

The primary data for species distribution analysis are species location records.  These localities for millions of plants and animal specimens are recorded in the world's natural history museums.  There is a growing recognition of the usefulness of these records for spatial modelling of biodiversity (Peterson and Stockwell 2002).  These data have also become easily accessible via the integration of the world's museum collection into a distributed database called the Species Analyst (Vieglais et.al. 1998). 

Developing maps of species distributions using multivariate models of species occurrence points with environmental variables is now a widespread practice in biodiversity science (eg. Scott \textit{et al.} 2002). The primary challenge of biodiversity mapping is to develop accurate maps based on statistical relationships rather than actual observations, particularly using \textit{ad hoc} museum collections data rather than controlled survey data. The Lifemapper team chose to use an evolutionary algorithm called the Genetic Algorithm for Rule-set Production (http://biodi.sdsc.edu), or GARP (Stockwell and Noble 1992, Stockwell and Peters 1999, Stockwell 1999).   In comparisons with other algorithms, the GARP algorithm has been shown to be the best available method for reliable species predictions using small sets of \textit{ad hoc} data typically returned from museum databases (Peterson and Stockwell 2002, Stockwell and Peterson 2002a, Stockwell and Peterson 2002b). The robustness of the algorithm has contributed to its use in projects requiring the development of large numbers of species distribution models, such as change in ecological communities due to climate change (Peterson \textit{et al.} 2002). One of the purposes of this paper is to discuss describe how the qualities of the GARP genetic algorithm contribute to the success of the project.

The second main challenge of developing a fauna and flora for the world is the computational scale of the project.  This problem is computationally intensive in many aspects:

\begin{itemize}
\item the number of species in the world that could be mapped, 9,600 of which are birds alone (Sibley and Monroe 1990),
\item the fine scale of maps that could be produced. e.g. a map of the world at a resolution of 1km at the equator has approximately 1,000,000,000 cells,  
\item the number of replicates that need to be produced for each species for estimating statistical variance,
\item the need for computing distributions under alternative scenarios such as climate change, and geographic invasion, and
\item the need to recompute the maps when new data becomes available.
\end{itemize}

The need to maintain an up-to-date resource by recomputing maps when new data becomes available ensures a perpetual demand for computing resources.  The combination of millions of idle computers around the world connected to the Internet forms the infrastructure for Internet grid computing which seeks to exploit otherwise idle workstations, PCs and bandwidth to create powerful distributed computing systems.  Internet computing was popularized by the SETI@home project which enlisted personal computers to analyse data for indications of extraterrestrial intelligence. As SETI@home is now running on half a million PCs and delivering 1,000 CPU years per day, it is currently the fastest (admittedly special purpose) computer in the world.  

The goal of the Lifemapper project is to become a major component in the biodiversity informatics infrastructure.  The archive of maps and models for the world's species developed through Lifemapper will be an invaluable resource for researchers in terrestrial, marine and freshwater environments.  This paper describes the Lifemapper resource with particular reference to the genetic modelling algorithm that serves as the core computational component of the screensaver modelling program.

\section{Methods}

The general approach to producing maps of species' distributions is to develop a multi-variate statistical model of known species' occurrence records and environmental variables. The species occurrence data are gathered from a number of biological collections housed at several museums and herbaria worldwide. Those institutions have their specimen databases linked and integrated through the Species Analyst (Vieglais \textit{et al.} 1998) project.  The environmental information is composed of a set of global geographic coverages, called environmental layers. Each layer represents one particular environmental parameter, such as temperature, rainfall, land use, elevation, among others. The layers are continuous grids, where each cell contains the value of an environmental parameter at a location. 

A model for estimating the probability of occurrence of a species is developed from the species occurrences and the environmental variables.  A map of the distribution of the species is then produced by using the values of the environmental variables at each grid cell to predict the probability of the occurrence of the species over the entire grid (eg. Fig. 1).  

\begin{figure}
\includegraphics{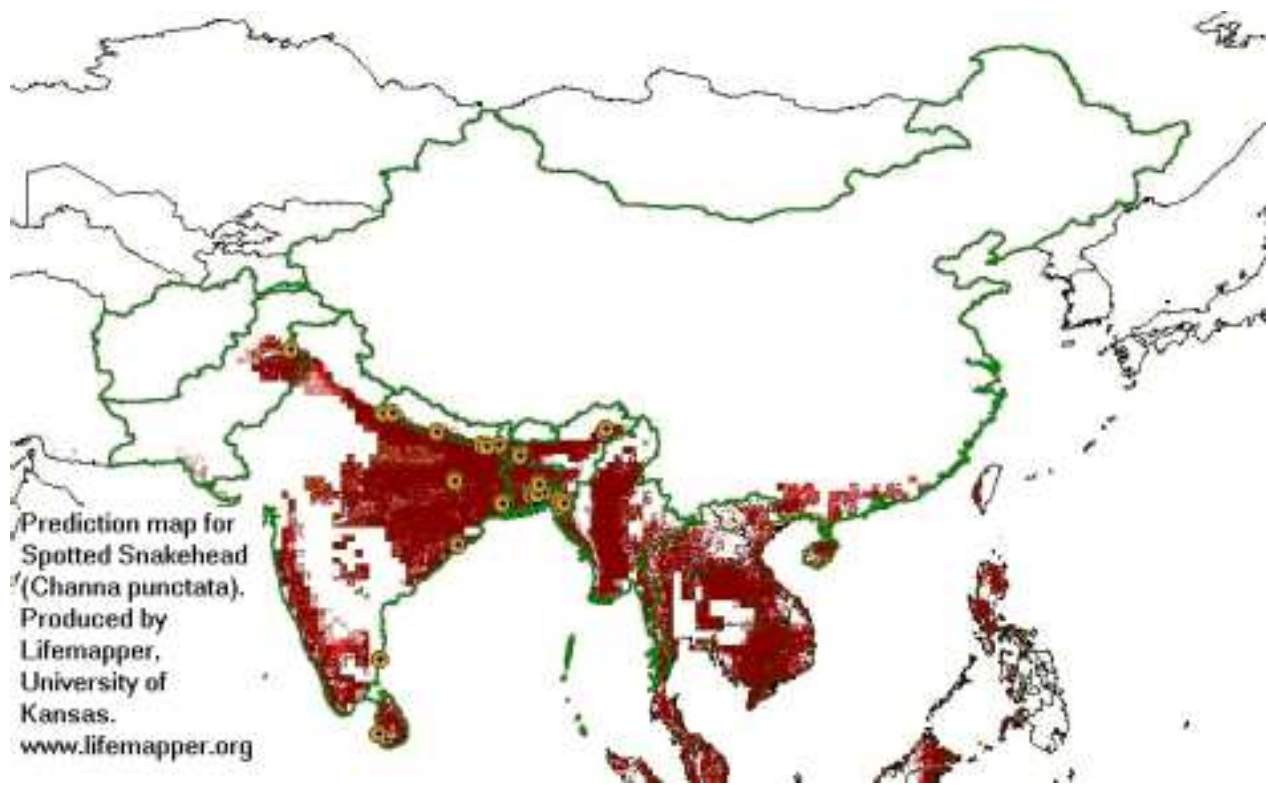}
\caption{
An example of the GARP predicted distribution of a species of fish,
the Spotted Snakehead. Areas predicted (shaded) in South-eastern Asia
where there are no data points (circles) have favourable
environmental conditions for the species.}
\end{figure}

The Lifemapper system is composed of components for harvesting species occurrence data from museum databases, developing models via the screensaver program, storing the results in an archive, and enabling access to information and membership management functions. The schematic diagram for Lifemapper is shown in Figure 2. This is described in three stages, preprocessing, processing and post-processing.

\begin{figure}
\includegraphics{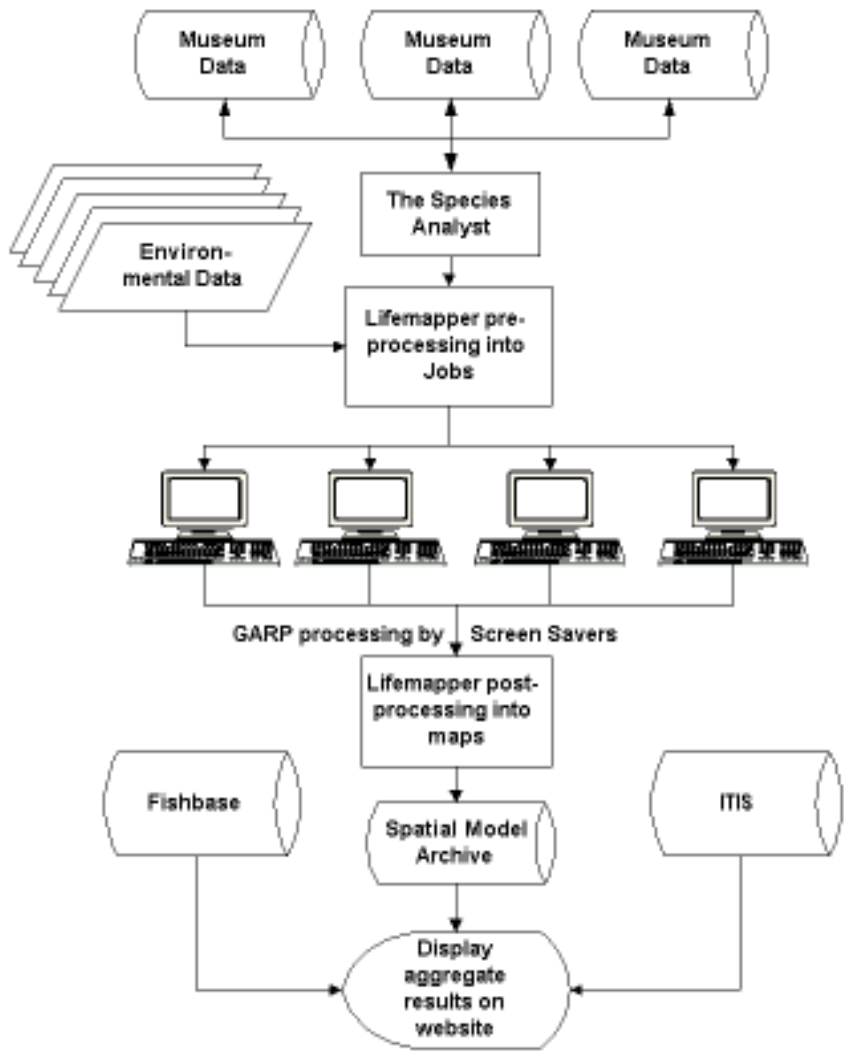}
\caption{The architecture of the Lifemapper system. Preprocessing consists of
data collation via distributed Internet query, processing via
distributed PC screensaver clients, and post-processing of results
into a digital library of spatial images and other information.}
\end{figure}

\subsection{Preprocessing}

The Lifemapper program retrieves georeferenced data from a network of biodiversity databases that are accessible over the Internet. At this time, almost all sources utilize the Species Analyst architecture, which uses Z3950 protocol and XML for direct information retrieval from participating databases over the Internet.   The first stage consists of periodically scanning through the data sources, building and updating the list of unique scientific names that are currently available for retrieval. The second stage builds a list of georeferenced records, using the list of names. 

The environmental layers consist of a set of geographical coverages that describe the main environmental parameters that may affect species' geographical distributions: temperature, precipitation, rainfall, solar radiation, terrain elevation and slope, tree coverage, among others.  

The original climate data used by Lifemapper was obtained from the Intergovernmental Panel on Climate Change (http://www.ipcc.ch/) at a scale of 0.5 degrees and processed to 1km resolution to match the other data sets. All variables are year averages from 1961 to 1990, of the averages for the months of January and July for the same period: cloud cover, diurnal temperature range, ground-frost frequency, maximum temperature, mean temperature, minimum temperature, precipitation, solar radiation, vapor pressure, wet-day frequency, and winds.  

The terrain related layers were obtained on a scale of 1:250.000. Those layers were later processed to grids with 1 km cells.  

\begin{itemize}

\item Aspect: describes the direction of maximum rate of change in the elevations between each cell and its eight neighbors. It can essentially be thought of as the slope direction; 

\item Flow directions: defines the direction of flow from each cell in the Digital Elevation Model to its steepest down-slope neighbor;

\item Flow accumulation defines the amount of upstream area draining into each cell. It is essentially a measure of the upstream catchment area. The flow direction layer is used to define which cells flow into the target cell

\item Slope: describes the maximum change in the elevations between each cell and its eight neighbors. The slope is expressed in integer degrees of slope between 0 and 90

\item Compound Topographic Index or CTI is commonly referred to as the Wetness Index is a function of the upstream contributing area and the slope of the landscape. The CTI is calculated using the flow accumulation (FA) layer along with the slope. In areas of no slope, a CTI value is obtained by substituting a slope of 0.001. This value is smaller than the smallest slope obtainable from a 1000 m data set with a 1m vertical resolution. 

\end{itemize}

Vegetation and land use data was obtained from University of Maryland at 1km resolution derived from remotely sensed Landsat NDVI data: percentage of tree cover, and use land cover type.

\subsection{Processing}

A genetic algorithm is a well known machine learning method (Holland 1975). This class of algorithms is inspired by the concept of evolution via natural selection and is based on the idea of evolving solutions to problems in a way which is analogous to the way organisms evolve.  The idea is to create a set of potential solutions to a problem (the population of organisms) and then iteratively modify and test this set until an optimal solution is found.

In GARP, a population is a set of individual rules for predicting the presence or absence of a species at a cell. Rules are composed of ``chromosomes'' that encode the coefficients for the variables in the model. For example, each individual in a population has chromosomes for climate, geology, aspect and the abundance of a species.  The ``fitness'' of each individual in a population is assessed at each iteration (``generation'') and determines the ``reproductive success'' of the individual. Fitness of an individual is based on the statistical significance of the rule in predicting presence and absence. A new population of candidate solutions is formed from the fittest individuals, with each individual altered from its parents by random mutation operators: point mutation, and crossover.  This procedure is repeated until a stopping criteria is met. GARP utilizes a rule archive which maintains a set of the best, uniquely different rules. The benefit of the archive is that the algorithm does not converge on a single rule, or solution, but develops a set of solutions that together produce a robust predictive model.

GARP uses four types of rules simultaneously in the population of individuals: atomic, Bioclim rules, range rules and Logit rules.

\begin{itemize}
\item An atomic rule uses only a single value of the variable in the precondition of the rule, e.g. "if the average annual temperature is 23 degrees C and the geology type is 4 then the species is present."

\item A Bioclim rule is based on the form of model used in the Bioclim program used by Nix (1986), for predicting the range of a species from their environmental tolerances. The Bioclim rule encloses the range of the environmental values of the data points in an ``envelope''. The distribution of the species is predicted at those points that fall within that environmental envelope, and absence predicted outside those points. For example, "If the annual average temperature is greater than 23 degrees C and less than or equal to 29 degrees C... predict presence".  The negation of a Bioclim rule can be used to predict presence or absence.

\item A range rule is a generalization of the Bioclim rule. In a range rule a number of variables may be regarded as irrelevant, that is, all possible predictor variables need not be used in the rule. When applying these rules, the undefined variables are inconsequential and may take on any value.  Range rules also allow negation, i.e. the rule applies outside of the indicated range.

\item Another family of rules called logit rules are useful when species respond to the environment through environmental gradients.  Logit rules are an adaptation of logistic regression models to rules.  For example the logistic regression gives the output probability \textit{p} that determines if a rule should be applied where \textit{p} is calculated using: $p = \frac{1}{e^{-y} + 1}$ and \textit{y} is the sum of the linear equation of the form: $a + a*a + b + b*b ... + n + n*n$ where $a...n$ represent variables and their coefficients. If \textit{p} is greater than 0.75 then the logit rule predicts presence.
\end{itemize}

Once the set of rules is obtained, GARP evaluates each rule at each cell to determine the presence or absence and applies the rule with the greatest expected accuracy. The resulting map is the prediction of the areas that would be the suitable for that species. The screen saver that runs on a Lifemapper user's computer produces one of these maps for every "job" that it computes  only one of many produced for a particular species.

\subsection{Post-processing}

A critical requirement of any grid application based on unreliable components is to ensure the trustworthiness of the returned calculations.  The system must ensure that faults in returned calculations do not affect the integrity of the overall calculation.  Sources of faults come from hardware and network sources, and users might provide a system to the grid without executing the application fully. 
 
Firstly, jobs for each species are sent out to many different processing sites.  When (and if) these jobs return, measures are in place to check the data conforms to the expected format, thus weeding out the dubious returns. Jobs will continue to be sent out for each species until sufficient returns are in place.

Secondly, the confidence in the calculation of the species distribution is increased by averaging the returned results.  Each map is combined with all the other maps that have been computed for that species, then the values are divided by the total number of models computed. This gives us a composite model that has values ranging from 0 to 100. So, if 3 maps predict presence in a particular area, and 1 map does not, the area has a 75 percent predicted presence for that species.

The results acquired from the screensaver processing units are stored in a database at the Lifemapper server site.  These include maps of species distribution and statistics on the identity of the processing unit, the data used to develop the map, and which museums contributed the data.  The Lifemapper database provides a resource for research scientists interested in species distributions, and raw data for producing reports on the creation of the overview for compiling membership statistics. Other data sources utilized by Lifemapper at this stage include the ISIS database for standard scientific nomenclature (http://sis.agr.gc.ca/pls/itisca) and FISHBASE  (http://www.fishbase.org) for a listing of common names that match our scientific names.

\subsection{Membership Management}
Lifemapper is a distributed computing project, which means that the participants (Lifemappers, as we call them) contribute by downloading our screensaver, doing number crunching, and enrolling new members. Members can form goups, whose members have common interests and goals. For example, a group of Kansas Bird watchers can meet each other and help the Lifemapper project out at the same time by creating a group. Once the group is created, it's attributes may be by species, by name or number of jobs.

The website displays progress reports on numbers of data and models developed. Taxa reports show statistics on numbers of online accessible records of unique biological names. Specimen reports deal with unique specimen records that can be accessed online. These numbers vary enormously between species and may increase over time.  The changes are tracked by unique taxon name additions, country from which most specimens came, and museum of origin.

\subsection{Archive Access}

Users can access the archive of maps of species distributions developed by Lifemapper by searching for a the species common or scientific name.  A query on 'Bobcat' for example give a listing of alternative possible species (\textit{Lynx rufus} or \textit{Lynx rufus rufus}) the number of contributing data sources, the number of georeferenced specimen records contributing to the model, and the number of GARP models computed. The maps can be viewed with a range of backgrounds, and at a range of scales.  There is provision for feedback or review of the quality of the map for each particular species.

Maps can also be accessed via the Lifemapper Web Mapping Service (WMS).  Directions are given on the web site for constructing URLs that return maps as an image in GIF format to a standard web browser.  The Lifemapper site also implement an OGC compliant web service that returns the capabilities of the site in XML format metadata - a machine and human readable description of the WMS's information and acceptable request parameters. This metadata allows a remote program to construct valid map requests.

\section{Discussion}

Reliability of prediction using data of variable quality is crucial to the success of analysis of \textit{ad hoc} data. The data from museum databases are collected in an \textit{ad hoc} manner and species are often represented by very few samples.  Thus, methods need to perform well on unstructured surveys and small samples.  In general, we want a method to be robust and efficient in its use of data. Although robustness is not easy to define, generally means it performs well over a wide range of conditions.  The capacity of genetic algorithms to perform will in poorly structured domains is well known (e.g. Goldberg 1989). The robustness of GARP has been confirmed in a number of recent studies evaluating the effects of biological data sample size and sampling bias, and variation in the types of responses of species to the environmental variables.

Due to the efforts of museums to digitize their data, the data available and numbers of species are increasing rapidly.  However, given the vast number of species (millions of insects alone) and the scant data on many of them, it is imperative that the modelling method is "data efficient" - i.e. produce accurate models using the minimum data.  Data efficiency can be measured by how few data are required to develop an accurate model. Evaluations of the GARP algorithm show an average of 90\% of maximum accuracy was achieved with a minimum of 10 data points (Stockwell and Peterson 2002b).  In this same study, the genetic algorithm half as many data points as logistic linear regression and fine resolution surrogate models.  Therefore GARP was approximately twice as data efficient at small sample sizes.  It also achieved accuracy equally or exceeding logistic regression and surrogate modelling methods on up to 100 data points.  The genetic algorithm and the rule-set composed of multiple models provides an accurate 'generalized' model covering a range of data sizes.  

This capacity to perform well over a range of data characteristics is also shown with respect to the environmental data.  Species typically have a unimodal response to the environment, i.e. the response surface describing the suitability of habitat will have an optimal value for a particular variable such as temperature, and fall to zero with increasing distance from that optimum.  However, other forms of response surface are possible.  In the case of surrogate models where environmental variables are based on vegetation types, optimal suitability is related to unique values or categories in a variable. The response therefore is highly non-linear, in fact, a discontinuous or piecewise distribution.  In order to make full use of the range of responses to environmental correlates we need a system that can recognize different types of correlation that are possible with different types of environmental data. The results of Stockwell and Peterson (2002b) show the capacity of the genetic algorithm and the rule-set it develops to provide an accurate model with a range of types of environmental variables. In contrast, the accuracy of stepwise additive logistic regression models decreased with the inclusion of a variable composed of categorical vegetation types. 

Finally, a pervasive problem in practical modelling is that of overfitting, or the production of models that are too specific for a given data set, and therefore perform poorly when tested on independent, new data.  The iterative approach of genetic algorithms can reduce overfitting by evaluating the fitness of models repeatedly on random samples of data.  The problem of overfitting is particularly severe where small data sets are not randomly selected - a situation known as bias.  Again, in a comparative study using controlled amounts of bias, the GARP genetic algorithm performed better than alternative approaches (Stockwell and Peterson 2002a).

These results provide comprehensive support for the use of the GARP algorithm in the development of this large information resource. While the results do not preclude the development of more robust methods, they show the importance of understanding the effects of data quality and evaluating robustness of prediction in alternative algorithms.  Expanding the conditions under which algorithms can perform reliably is crucial for expanding the applicability of models to readily available data sources. The advantages of the genetic algorithm in GARP rely heavily on the use of machine learning methods, which along with neural nets, decision trees, and adaptive agent-based learning methods are increasing being used in ecological applications (Fielding 1999). As a result of this work, users can have greater confidence in Lifemapper's results for application to biodiversity research, education and conservation worldwide.  The potential uses are many:

\begin{itemize}
\item Researchers will be able to model and simulate the spread of emerging diseases, plant and animal pests, or invasive species of plants and animals and their effects on natural resources, agricultural crops and human populations.

\item Environmental scientists will be able to model and predict the effects of local, regional or global climate change on Earth's species of plants and animals.

\item Land planners and policy makers will be able to identify the highest priority areas for biodiversity conservation.

\item Teachers, students and the public will be able to discover and map their backyard biodiversity and how it might be affected by changes in rainfall or temperature or by the spread of other species. 
\end{itemize}

Biodiversity science has yet to develop a complete electronic atlas of the Earth's biological diversity. Lifemapper will help science to achieve this goal, by developing an ever increasing flora and fauna of the world's taxa as data becomes available.  The number of taxa in Lifemapper has been increasing explosively as more databases come online, currently greater than 100,000 species (Figure 3).  The project will assist in the use of this knowledge to better understand and conserve Earth's biological diversity, and inform environmental solutions for Earth's biological diversity.
\begin{figure}
\includegraphics{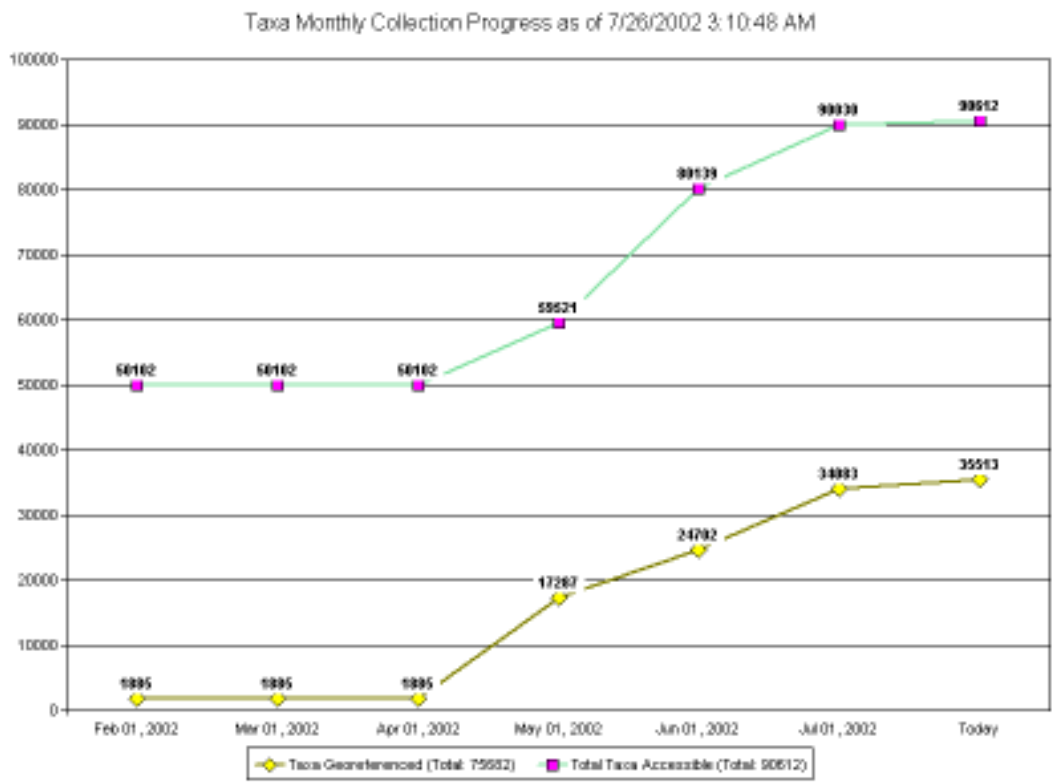}
\caption{The increase in the numbers of taxa modeled by GARP in the Lifmapper
system.}
\end{figure}

We consider the project will provide incentive for the museum collection community to database, clean and contribute their data to biodiversity databases such as the Species Analyst (Vieglais \textit{et al.} 1998).  As well as showing the scrolling display of the list of contributing museums on the screensaver, the system also provides detailed feedback to museum data providers on how many times their data points have been viewed in maps via collection analysis statistics.  More detailed feedback including possible errors in the data will be provided in the future.  
This relationship with the museums increases the value of participation and helps them see benefit in contributing free access to data for use by the project.  The success of this strategy can be seen in the growth in number of records provided by contributing museums for use in the project, currently greater than 300,000 records (Figure 4).

\begin{figure}
\includegraphics{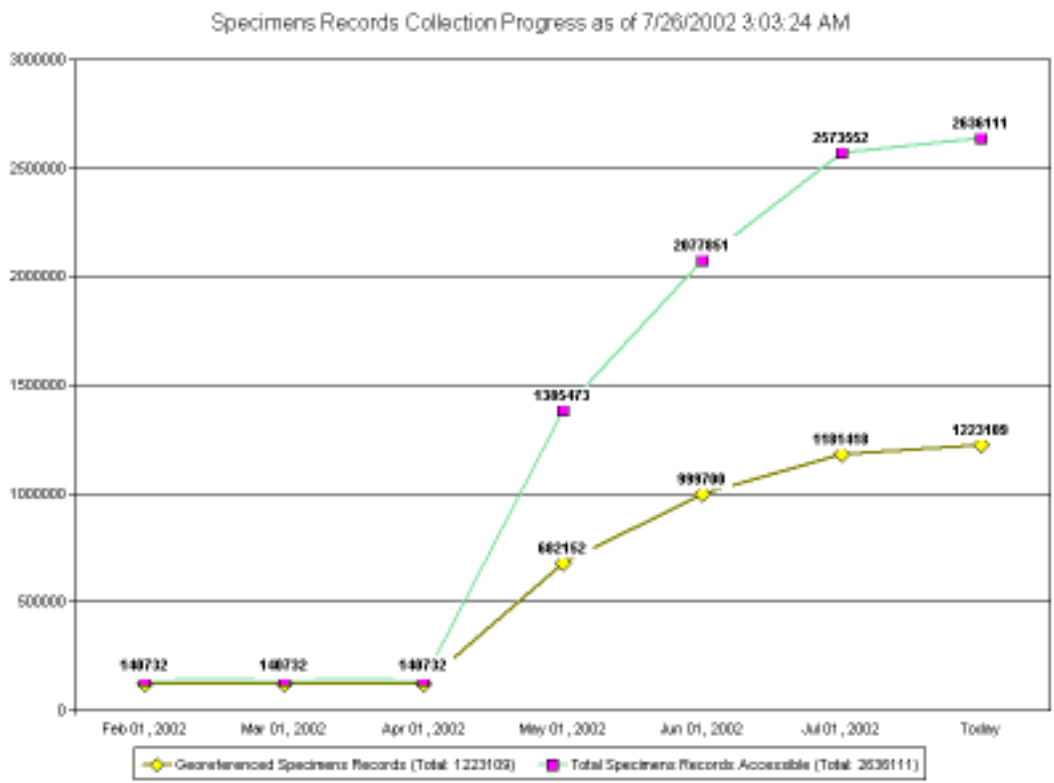}
\caption{
The increase in the numbers of records available to the Lifemapper
system from participating museums.}
\end{figure}

Another important benefit of Internet computing is to encourage the involvement of the general community in science.  We expect that Lifemapper will be an important tool for outreach to the general community, not only through the screen saver program, but also through a spatial query interface addressing the question ``What grows/prowls in my backyard/county?''.

Lifemapper is an example of a real world application of genetic algorithms coupled with the computational resources of Internet computing.  It will require ongoing use of large computing resources.

As such it demonstrates a dimension of problem on a par with bioinformatics, medical and other applications of large scale evolutionary algorithms.  Lifemapper demonstrates that biodiversity researchers can cast their problems in a form suitable for execution on home computers and then persuade the public that their problem is important enough to justify the expenditure of ``free'' cycles.  In the future, Lifemapper will be an ongoing computational resource for addressing a wide range of computationally intensive biodiversity research projects, and an indispensable component of the informatics infrastructure of biodiversity science.

\section{Acknowledgments}

We thank the National Science Foundation for its support via a grant from the Division of Environmental Biology (DBI9873021).

\end{document}